\documentclass{PoS}

\usepackage{amsmath,amssymb}

\newcommand{\bea}{\begin{eqnarray}}
\newcommand{\beas}{\begin{eqnarray*}}
\newcommand{\nn}{\nonumber}
\newcommand{\eea}{\end{eqnarray}}
\newcommand{\eeas}{\end{eqnarray*}}
\newcommand{\bd}{\begin{displaymath}}
\newcommand{\ed}{\end{displaymath}}
\newcommand{\be}{\begin{equation}}
\newcommand{\ee}{\end{equation}}
\newcommand{\bi}{\begin{itemize}}
\newcommand{\ei}{\end{itemize}}

\newcommand{\as}{\alpha_s}

\newcommand{\Z}{\textbf{\textit{Z}}}
\newcommand{\T}{\textbf{\textit{T}}}

\newcommand{\captionfonts}{\small}
\makeatletter  
\long\def\@makecaption#1#2{%
  \vskip\abovecaptionskip
  \sbox\@tempboxa{{\captionfonts #1: #2}}%
  \ifdim \wd\@tempboxa >\hsize
    {\captionfonts #1: #2\par}
  \else
    \hbox to\hsize{\hfil\box\@tempboxa\hfil}%
  \fi
  \vskip\belowcaptionskip}
\makeatother   

\title{Analysis of the anomalous-dimension matrix of $n$-jet operators at 4 loops}

\ShortTitle{Anomalous-dimension matrix of $n$-jet operators}

\author{\speaker{Leonardo Vernazza}
        \thanks{AvH fellow.}\\
        {\sl
Institut f\"ur Physik (THEP),\\ Johannes Gutenberg-Universit\"at, D--55099 Mainz, Germany}\\
        E-mail: \email{vernazza@uni-mainz.de}}

\abstract{Recently, an all-order conjecture for the anomalous-dimension
matrix of $n$-jet operators in SCET was proposed, which allows one to
predict the structure of the infrared divergences of dimensionally
regularized, massless gauge-theory scattering amplitudes with an
arbitrary number of legs and loops. The conjecture is severely
constrained by soft-collinear factorization, non-abelian
exponentiation, and the behavior of amplitudes in collinear limits.
Using these constraints, a diagrammatic analysis has shown that the
anomalous dimension involves only two-parton correlators up to three
loop order. The only exception is given by a single color structure
multiplying a function of conformal cross ratios depending on the
momenta of four external partons, which would have to vanish in all
two-particle collinear limits. We extend this analysis by completing
the diagrammatic analysis at four loop, and we find that additional
functions which vanish in all two-particle collinear limits may
arise.}

\FullConference{The 2011 Europhysics Conference on High Energy Physics-HEP 2011,\\
		July 21-27, 2011\\
		Grenoble, Rh\^one-Alpes, France}

\begin{document}

\section{Infrared divergences of massless gauge-theory
scattering amplitudes}

The problem of predicting the structure of infrared
singularities of on-shell $n$-particles scattering
amplitudes in massless QCD simplifies, if one realizes
that they can be put in one-to-one correspondence with
UV divergences of operators defined in soft-collinear
effective field theory (SCET)
\cite{Becher:2009cu,Gardi:2009qi,Becher:2009qa}.

This idea implies that infrared divergences can
be studied  by means of standard renormalization-group
techniques. The IR divergences of $n$-point scattering
amplitudes can be absorbed into a multiplicative
renormalization factor $\Z$, which is related to an
anomalous dimension \textbf{$\Gamma$}. The $\Z$-factor
and the anomalous dimension \textbf{$\Gamma$} are
matrices in color space, i.e. they mix amplitudes with
the same particle content but different color
structure.

The predictive power of this approach relies on
the fact that the anomalous dimension \textbf{$\Gamma$}
is strictly constrained by the structure of the effective
field theory, as well as by other elements, like the
non-abelian exponentiation theorem and the two parton
collinear limit. In
\cite{Becher:2009cu,Gardi:2009qi,Becher:2009qa}
a form of the anomalous dimension \textbf{$\Gamma$}
compatible with all these constraints has been derived,
whose central feature is that only pairwise correlations
among charges and momenta of different partons are
allowed. Using the color-space formalism
it explicitly reads
\be\label{1}
\boldmath{\Gamma}(\{\underline p\},\mu) =
\sum_{(i,j)} \frac{\T_i\cdot\T_j}{2} \gamma_{\rm cusp}(\as)
\,\ln\frac{\mu^2}{-s_{ij}} + \sum_i \gamma^{i}(\as),
\ee
where $s_{ij} \equiv 2\sigma_{ij}p_i\cdot p_j+i0$,
and the sign factor $\sigma_{ij} = +1$ if the
momenta $p_i$ and $p_j$ are both incoming or outgoing,
and $\sigma_{ij} = -1$ otherwise. The sum runs over the
$n$ external partons, and we refer to \cite{Becher:2009qa}
for further detail.

One of the major statements of
\cite{Becher:2009cu,Gardi:2009qi,Becher:2009qa}
is that the conjecture (\ref{1}) should hold to 
all orders in perturbation theory. If true, this is an
intriguing result, because it implies a semi-classical
origin of IR singularities, and it allows to shed new
light on the deeper structure of the strong interaction.
For instance, an interesting consequence
is the prediction that the cusp anomalous
dimension of quarks and gluons should be equal to
the quadratic Casimir operator $C_{\rm R}$ in the
fundamental or adjoint representation times a
universal coefficients. In other words, the cusp 
anomalous dimension of quarks and gluons should 
obey Casimir scaling to all orders in perturbation 
theory. Such a prediction is highly non-trivial, because 
it is expected not to hold anymore at the non-perturbative
level, which is supported by recent investigations based
on the AdS/CFT correspondence 
\cite{Armoni:2006ux,Alday:2007hr,Alday:2007mf}.

The validity of (\ref{1}) was studied explicitly
to three-loop order by means of  a diagrammatic
analysis, \cite{Becher:2009qa,Dixon:2009ur}.
It was found that only one new color
structure can arise, at three loop,
which has to depend on the momenta of four external
partons. In order to be consistent with all
the constraints, however, the corresponding
coefficient function, which encode the momentum
dependence, must be highly nontrivial, as it must
vanish in all collinear limits. Explicitly
the additional term reads
\be\label{2}
\boldmath{\Delta \Gamma}(\{\underline p\},\mu) =
\sum_{(i,j,k,l)} f^{ade}f^{bce} \T_i^a \T_j^b \T_k^c \T_l^d
\;F(\beta_{ijkl},\beta_{iklj}-\beta_{iljk}),
\ee
where $\beta_{iklj}= \ln
\frac{(-s_{ij})(-s_{kl})}{(-s_{ik})(-s_{jl})}$ is the
logarithm of the conformal cross ratio. An
example of such a function \cite{Dixon:2009ur}
is $F(x,y) = x^3(x^2-y^2)$, but it was
recently excluded using additional constraints
obtained exploiting the high-energy Regge limit
of the anomalous dimension \textbf{$\Gamma$}
\cite{DelDuca:2011xm,DelDuca:2011ae}.

In this talk we report about work in progress
\cite{Mainz2}. We extend the diagrammatic analysis
\cite{Becher:2009qa} to four loops in perturbation
theory, in order to look for new color structures
which could arise and break the simple structure of
the anomalous dimension (\ref{1}). In particular,
we look for new terms which could lead to a
violation of the Casimir scaling of the cusp
anomalous dimension of quarks and gluons.
We refer to \cite{Mainz2} for further details.

\section{Diagrammatic analysis at four loops}

\begin{figure}[t]
\begin{center}
  \includegraphics[width=0.90\textwidth]{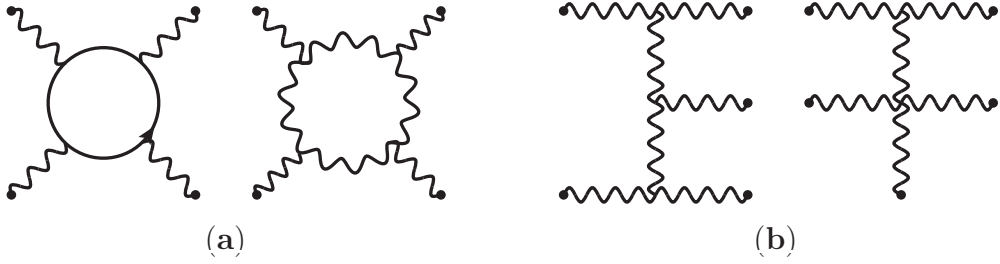}
\end{center}
  \caption{Gluon webs contributing to the soft
  anomalous dimension matrix at four loops.}
  \label{soft}
\end{figure}

The correspondence between infrared singularities
of gauge scattering amplitudes and UV poles of matrix
elements of SCET operators can be seen by considering
off-shell $n$-parton Green's function with
large momentum transfer $s_{ij}=(p_i\pm p_j)^2$ and
small off-shellness $p_i^2 \ll s_{ij}$. In the effective
theory these Green's functions are represented by the
matrix element of UV renormalized $n$-jet operators:
\be\label{3}
{\mathcal G}_n(\{p\}) = \lim_{\epsilon\to0}\sum_{i}
{\cal C}_{n,i} (\mu)\langle{\cal O}_{n,i}^{\rm ren} (\mu)\rangle
 = \lim_{\epsilon\to0}\sum_{i,j}
{\cal C}_{n,i}(\mu)Z_{ij}(\mu,\epsilon)
\langle{\cal O}_{n,j}^{\rm bare} (\epsilon)\rangle.
\ee
In the second identity we have written explicitly
the renormalization factor $\Z$ of the bare SCET operators.
To obtain on-shell $n$-parton scattering amplitudes
one takes the limit $p_i^2\to 0$. In this way, on the one
hand infrared divergences are introduced in the Green's
function, which are regularized in $d=4-2\epsilon$. On the
other, the matrix elements of the bare operators become trivial,
because the collinear and soft scale are set to zero, and
all loops integrals in the effective theory become scaleless
and vanish. The $n$-parton scattering amplitude free of
infrared divergences is therefore equal to the
Wilson coefficient $\mathcal{C}_{n}$,
\be\label{4}
\mathcal{C}_{n,i}(\{p\},\mu) = \lim_{\epsilon\to0}\sum_{i,j}
\left(Z^{-1}\right)_{ij} {\mathcal G}_{n,j}(\epsilon\{p\}),
\ee
times trivial color and Dirac structures from the
operator matrix element.
The logarithm of the renormalization factor $\Z$
is related via $\Gamma = -d\ln \Z / d\ln \mu$ to the
anomalous dimension matrix $\Gamma$ governing the RG
evolution equation of the $n$-jet SCET operators
${\cal O}_n^{\rm ren}$. $\Gamma = \Gamma_{c+s}$
is determined by the collinear and soft modes, and
collinear-soft factorization assures that
$\Gamma_{c+s} = \Gamma_s + \sum_i\Gamma_c^i$.
The collinear anomalous dimension is known, 
$\Gamma_c^i= -\Gamma^i_{\rm cusp} L_i + \gamma_c^i$,
where $L_i  = \ln \mu^2/p_i^2$ is a collinear logarithm, 
therefore the conjecture (\ref{1})
becomes a prediction for $\Gamma_s$, with the
constraint $\partial \Gamma_s (\mu)/ \partial L_i
= \Gamma_i^{\rm cusp}$.
This important constraint originates from the
requirement that collinear logarithms must cancel
between  $\Gamma_s$ and $\sum_i\Gamma_c^i$,
in order to match the dependence on the hard
scale alone in $\Gamma=\Gamma_{c+s}$.
As a consequence, the conjecture (\ref{1})
can be proven order by order based on a diagrammatic
analysis of the anomalous dimension of the soft
function. The latter is given by the vacuum
expectation value of $n$ soft Wilson lines.
The non-abelian exponentiation theorem guarantees
that the corresponding anomalous dimension receives
contributions only from single-connected gluon
webs attached to the $n$ Wilson lines of the
soft operator. An additional constraint comes
from the two-parton collinear limit: when two
partons become collinear, a $n$-parton amplitude
splits into a $(n-1)$-amplitude times a splitting
function. One finds that the anomalous dimension
of the splitting function can only depend on the
color and momenta of the two collinear partons.

Using all these results, new structures
contributing to the anomalous dimension can
arise at four loops, which are related to the webs
in Fig. \ref{soft}. The webs in (a) are
proportional to higher Casimir invariants.
The possibility of
having new structures of this type, linear
in the cusp angle,
$\beta_{ij} \equiv \ln \mu^2/p_i^2
+ \ln \mu^2/p_j^2 - \ln \mu^2/(-s_{ij})$,
has been already excluded \cite{Becher:2009qa}. 
Here we focus
on structures which have a different
momentum dependence, such as a dependence
on the logarithm of the conformal cross
ratio defined above. We find that the
following structures are compatible
with constraints from the
soft-collinear factorization:
\be\label{5}
{\Delta \Gamma} = \sum_{(i,j)}
\left[{\cal D}_{iijj}\,g_{4}(\as) +
{\cal D}_{iiii}\,g_{5}(\as)\right] +
\sum_{(i,j,k,l)}{\cal D}_{ijkl}\,
G_1(\beta_{ijkl},\beta_{iklj}-\beta_{iljk}),
\ee
where $G_1(x,y) = G_1(-x,y)$ has to be even 
in its first argument to match the symmetries 
of ${\cal D}_{ijkl} = d_F^{abcd}\left(\T^a_i \T^b_j \T^c_k \T^d_l\right)_+$. 
Out of the terms proportional 
to $g_4(\as)$ and $g_5(\as)$, only the latter
is compatible  with the collinear constraint. 
However, since
it does not have an explicit scale dependence, 
it would not affect the Casimir scaling of the 
cusp anomalous dimension. 
Considering then the 
term proportional to $G_1(x,y)$, we 
see that, similarly to the term proportional to 
$F(x,y)$ in (\ref{2}), its presence is compatible 
with the collinear limit only if $G_1(x,y)$ 
vanishes in all two-particles collinear limits. 
An explicit example is $G_1(x,y) = x^2(x^2-y^2)^2$,
as found in \cite{DelDuca:2011xm} as well.
However, we find that this example, too,
is ruled out by the constraint from Reggeization. 
In the high energy limit, in fact,
$G_1(x,y) = x^2(x^2-y^2)^2$ contains 
terms proportional to $\ln (s/(-t))^4$. 
If present, these leading logarithms (with the corresponding 
nontrivial color structure) would appear 
as an additional contribution to the 
Regge trajectories of gluons, 
which are known to NLL \cite{Fadin:1995xg,Fadin:1995km,Fadin:1996tb,DelDuca:2001gu,Fadin:2006bj}, 
and those logarithms do not appear. Therefore, 
consistency with the Regge limit 
rules out $G_1(x,y) = x^2(x^2-y^2)^2$.
This of course does not exclude the 
existence of more complicated $G_1(x,y)$, 
in which the leading and the next-to-leading 
logarithms  cancel. 

The webs in Fig.\ref{soft}(b) involve up to 
five external partons, and have the following
color structure:
\be\label{6}
{\cal T}_{ijklm} =-{\cal T}_{jilkm}
                 =-{\cal T}_{klijm}
                 =-{\cal T}_{ikjlm}
                 =-{\cal T}_{ljkim}
= f^{adx}f^{bcy}f^{exy}(\T^a_i\T^b_j\T^c_k\T^d_l\T^e_m)_+ \;.
\ee
We consider the contributions to
the anomalous dimension compatible with the
symmetries of this object, and we look for 
those terms which satisfy soft-collinear 
factorization as well. We find:
\bea\label{7}\nn
{\Delta \Gamma} &=& \sum_{(i,j,k)} {\cal T}_{iijkk}\,\big[
\bar{g}_1(\as)\,\beta_{ij} + \bar{g}_2(\as)\,\beta_{jk} \big] \\
&&\hspace*{-1.0cm}+\,\sum_{(i,j,k,l,m)} {\cal T}_{ijklm}\,
G_2(\beta_{ijkm}, \beta_{ikmj}-\beta_{imjk},
\beta_{ijml}, \beta_{imlj}-\beta_{iljm}) \;.
\eea
Consistency with the collinear limit
eliminates the terms proportional to
$\bar g_1$ and $\bar g_2$. The last
term involving a nontrivial function of
the logarithms of the conformal cross ratios
is possible only if the function $G_2$
vanishes in all the collinear limits.
Its presence would however not affect the
Casimir scaling of the cusp anomalous
dimension.

In conclusion, we have analyzed possible new
contributions to the soft anomalous dimension
of $n$-jet operators in SCET at four loop in
perturbation theory. We find that
the only new color structures allowed 
would involve
four partons, with momentum dependence 
encoded in functions of the logarithm
of the conformal cross ratio. 
Those terms, however, would not affect the Casimir
scaling of the cusp anomalous dimension of
quarks and gluons.

\subsubsection*{Aknowledgments}

I thank the organizers of the conference for 
an exciting conference in a pleasant atmosphere, 
Li Lin Yang for comments on the manuscript and 
the Humboldt foundation for support.


\begin{thebibliography}{99}

\bibitem{Becher:2009cu}
  T.~Becher, M.~Neubert,
  Phys.\ Rev.\ Lett.\  {\bf 102 } (2009)  162001.
  [arXiv:0901.0722 [hep-ph]].

\bibitem{Gardi:2009qi}
  E.~Gardi, L.~Magnea,
  JHEP {\bf 0903 } (2009)  079.
  [arXiv:0901.1091 [hep-ph]].

\bibitem{Becher:2009qa}
  T.~Becher, M.~Neubert,
  JHEP {\bf 0906 } (2009)  081.
  [arXiv:0903.1126 [hep-ph]].

\bibitem{Dixon:2009ur}
  L.~J.~Dixon, E.~Gardi, L.~Magnea,
  JHEP {\bf 1002 } (2010)  081.
  [arXiv:0910.3653 [hep-ph]].

\bibitem{DelDuca:2011xm}
  V.~Del Duca, C.~Duhr, E.~Gardi, L.~Magnea, C.~D.~White,
  [arXiv:1108.5947 [hep-ph]].

\bibitem{DelDuca:2011ae}
  V.~Del Duca, C.~Duhr, E.~Gardi, L.~Magnea, C.~D.~White,
  [arXiv:1109.3581 [hep-ph]].

\bibitem{Armoni:2006ux}
  A.~Armoni,
  JHEP {\bf 0611 } (2006)  009.
  [hep-th/0608026].

\bibitem{Alday:2007hr}
  L.~F.~Alday, J.~M.~Maldacena,
  JHEP {\bf 0706 } (2007)  064.
  [arXiv:0705.0303 [hep-th]].

\bibitem{Alday:2007mf}
  L.~F.~Alday, J.~M.~Maldacena,
  JHEP {\bf 0711 } (2007)  019.
  [arXiv:0708.0672 [hep-th]].

\bibitem{Fadin:1995xg}
  V.~S.~Fadin, M.~I.~Kotsky and R.~Fiore,
  Phys.\ Lett.\ B {\bf 359} (1995) 181.

\bibitem{Fadin:1995km}
  V.~S.~Fadin, R.~Fiore and A.~Quartarolo,
  Phys.\ Rev.\ D {\bf 53} (1996) 2729
  [hep-ph/9506432].

\bibitem{Fadin:1996tb}
  V.~S.~Fadin, R.~Fiore and M.~I.~Kotsky,
  Phys.\ Lett.\ B {\bf 387} (1996) 593
  [hep-ph/9605357].

\bibitem{DelDuca:2001gu}
  V.~Del Duca and E.~W.~N.~Glover,
  JHEP {\bf 0110} (2001) 035
  [hep-ph/0109028].

\bibitem{Fadin:2006bj}
  V.~S.~Fadin, R.~Fiore, M.~G.~Kozlov and A.~V.~Reznichenko,
  Phys.\ Lett.\ B {\bf 639} (2006) 74
  [hep-ph/0602006].

\bibitem{Mainz2}
 V.~Ahrens, M.~Neubert and L.~Vernazza, in preparation.

\end{thebibliography}
\end{document}